\documentclass[11pt]{article}
\usepackage{amsmath,amssymb,color,graphics,epsfig,cite}

\usepackage{amsfonts}

\newcommand{\be}{\begin{equation}}
\newcommand{\ee}{\end{equation}}
\newcommand{\bea}{\setlength\arraycolsep{2pt} \begin{eqnarray}}
\newcommand{\eea}{\end{eqnarray}}
\newcommand{\nn}{\nonumber}

\def\ft#1#2{{\textstyle{\frac{\scriptstyle #1}{\scriptstyle #2} } }}
\def\fft#1#2{{\frac{#1}{#2}}}

\def\0{{\sst{(0)}}}
\def\1{{\sst{(1)}}}
\def\2{{\sst{(2)}}}
\def\3{{\sst{(3)}}}
\def\4{{\sst{(4)}}}
\def\5{{\sst{(5)}}}
\def\6{{\sst{(6)}}}
\def\7{{\sst{(7)}}}
\def\8{{\sst{(8)}}}
\def\sst#1{{\scriptscriptstyle #1}}

\def\del{{\partial}}

\begin{document}

\begin{center}
{\Large {\bf Large-Charge Limit of AdS Boson Stars \\ with Mixed Boundary Conditions}}

\vspace{20pt}

{\large Shi-Fa Guo, Hai-Shan Liu, H. L\"u and Yi Pang}

\vspace{10pt}

{\it  Center for Joint Quantum Studies and Department of Physics,\\
School of Science, Tianjin University, Tianjin 300350, China}

\vspace{40pt}

\underline{ABSTRACT}
\end{center}

It was recently shown that charged AdS boson stars can reproduce the universal structure of the lowest scaling dimension in the subsector of a CFT with fixed large global $U(1)$ charge $Q$. Using the model consisting of Einstein-Maxwell gravity with a negative cosmological constant, coupled to a $U(1)$-charged conformally massless scalar with the fourth-order self interaction, we construct a class of charged AdS boson star solutions in the large $Q$ limit, where the scalar field obeys a mixed boundary condition, parameterized by $k$ that interpolates between the Neumann and Dirichlet boundary conditions corresponding to $k=0$ and $\infty$ respectively. By varying $k$, we numerically read off the $k$ dependence of the leading coefficient $c_{3/2}(k) \equiv \lim_{Q\rightarrow \infty} M/Q^{3/2}$.  We find that $c_{3/2}(k)$ is a monotonously increasing function which grows linearly when $k$ is sufficiently small. When $k\rightarrow \infty$, $c_{3/2}(k)$ approaches the maximal value at a decreasing rate given by $k^{-3/2}$ . We also obtain a close form expression that fits the numerical data for the entire range of $k$ within $10^{-4}$ accuracy.

\vfill{\footnotesize shifaguo97@gmail.com \ \ \  hsliu.zju@gmail.com \ \ \ mrhonglu@gmail.com \ \ \ pangyi1@tju.edu.cn }


\thispagestyle{empty}
\pagebreak


\newpage

\section{Introduction}

The original AdS/CFT correspondence involves an exact duality between weakly coupled string theory in an asymptotically anti-de Sitter (AdS) spacetime and a conformal field theory (CFT) realised by brane configurations underlying the same AdS background \cite{Maldacena:1997re}. Subsequent studies generalized the AdS/CFT correspondence to gauge/gravity duality in which the gauge theory and the gravitational system need not to be strictly equivalent. Instead one focuses on certain universal features that appear in one side of the duality and tries to recover similar features in the other side. Recently, it was shown that the lowest scaling dimension in the subsector of a $D=3$ CFT with fixed large global $U(1)$ charge $Q$ takes the universal form \cite{Orlando1,Orlando2,Orlando3,Monin:2016jmo,Loukas:2018zjh,delaFuente:2018qwv,Badel:2019khk}
\be
\label{mq1}
\Delta(Q)=c_{3/2}\, Q^{3/2}+c_{1/2}\, Q^{1/2}+c_0+ c_{-1/2}\, Q^{-1/2} + {\cal O}(Q^{-3/2})\,.
\ee
(see also \cite{Gaume:2020bmp} for a recent review on this subject.) Although the result above was explicitly worked out for the weakly coupled $O(N)$ vector model \cite{Orlando1,Orlando2}, $SU(N)$ matrix model \cite{Orlando3} and $\mathbb{CP}^{N-1}$ model \cite{delaFuente:2018qwv}, simple argument based on dimension analysis indicates the relation \eqref{mq1} also applies to a large class of conformal fixed points of strongly coupled gauge theory \cite{Jafferis:2017zna}. Thus it becomes natural to reinterpret the universal CFT result from the gravity side.  This task has been carried out in \cite{delaFuente:2020yua,Liu:2020uaz} where the gravity dual of the ground state in the fixed global $U(1)$ charge sector of the $D=3$ CFT was identified with the global AdS boson star solutions in $D=4$ Einstein-Maxwell theory coupled to a charged massive scalar field. In the large charge limit, the energy of the boson stars reproduces the same behavior as in \eqref{mq1}. Interestingly, the leading power $3/2$ that appears in the large charge expansion of the boson star energy can also be simply explained by dimension analysis \cite{Gentle:2011kv}. Thus it is not surprising that gravitational result should match with its CFT counterpart in structure. The main merit of the holographic approach is the extraction of the coefficients $c_{3/2}\,,c_{1/2}\,,\cdots$ from the gravity solutions. Upon applying the holographic dictionary, one can read off the dependence of the coefficients on the parameters of the dual strongly coupled gauge theory which may turn out to be universal. In the previous work \cite{Liu:2020uaz}, we find that for holographic CFTs, the coefficient $c_0$ is of the order ${\cal O}(c_T)$, where $c_T$ is the central charge of the CFT, differing from the ${\cal O}(1)$ result obtained for weakly coupled non-gauge theory CFTs.

In this work, we would like to understand from the gravity side, how these coefficients vary when the dual CFT is deformed by multi-trace operators. In this case, the relevant solutions should be the AdS$_4$ boson stars with mixed boundary condition, thus differing from previous work in literature,{\it e.g.}~\cite{delaFuente:2020yua,Liu:2020uaz,
Gentle:2011kv,Astefanesei:2003qy,Hu:2012dx,Hartmann:2012gw,Buchel:2013uba,Dias:2016pma,Arias:2016aig,rotatingbs1}. Neutral AdS$_4$ solitonic solutions with mixed boundary condition were also studied in the context of designer gravity \cite{Hertog:2004ns}. Exact solutions of black holes with mixed scalar boundary condition are rather limited \cite{Lu:2013ura}. In the gravity setup, we shall consider the simple model
\be
S=\frac{1}{16\pi G}\int d^4x\sqrt{-g}\Big(R - \ft14 F^2 +\fft{6}{\ell^2}- (D_\mu \Phi) (D^\mu \Phi)^* - V(\Phi\Phi^*)\Big)\label{genlag}\,,
\ee
where $\ell$ is the vacuum AdS radius associated with the negative cosmological constant $\Lambda=-3/\ell^2$ and $\Phi$ is the $U(1)$-charged complex scalar whose covariant derivative and potential are
\be
D_\mu \Phi=\partial_\mu\Phi-{\rm i}q \ell^{-1} A_\mu\, \Phi\,,\qquad V(\Phi\Phi^*)=m^2 \ell^{-2}\Phi\Phi^*+\ft12\lambda \ell^{-2} (\Phi\Phi^*)^2\,.
\ee
In our convention, $m, q, \lambda$ are dimensionless parameters. Hereafter, we focus on $m^2=-2$, for which the scalar is conformally massless.  The theory then involves two nontrivial parameters $(q,\lambda)$. In the standard global AdS coordinates, near the AdS$_4$ boundary at $r=\infty$, the conformally coupled scalar field admits the Fefferman-Graham expansion
\be
\Phi=\frac{\phi_1(\vec{x})}r+\frac{\phi_2(\vec{x})}{r^2}+\cdots\,.
\ee
where $\vec{x}$ denotes the coordinates on the conformal boundary.
The renormalisable boundary condition on $\Phi$ that also preserves conformal symmetry takes the form
\be
\label{bc1}
\phi_2=k\phi_1^2\,.
\ee
The two special cases $k=0\,,\infty$ correspond to the standard Neumann ($\phi_2=0$) and Dirichlet ($\phi_1=0$) boundary conditions respectively. In the former choice, the scalar $\Phi$ is dual to a single-trace operator ${\cal O}_1$ of scaling dimension $\Delta=1$, with $\phi_1$ identified as its vaccum expectation value (VEV) and $\phi_2$ related to its source. In the latter, the scalar $\Phi$ is dual to another single-trace operator ${\cal O}_2$ of scaling dimension $\Delta=2$, and the role of $\phi_1$ and $\phi_2$ are interchanged. The two different boundary conditions correspond to two quantization schemes of the dual field theory that are related by a Legendre
transformation. Suppose we adopt the first method of quantization and insert a triple-trace deformation of the form $W=\ft{k}3{\cal O}_1^3$ for generic values of $k$. Such deformation in the CFT side is precisely captured by the mixing boundary condition \eqref{bc1} in the gravity side \cite{Witten:2001ua}. Our choice of the mixed boundary condition is different from the one adopted in \cite{Gentle:2011kv} where $\phi_2$ and $\phi_1$ obeys a linear relation that breaks the conformal symmetry.

In this paper, we construct AdS boson stars in theory (\ref{genlag}) with the mixed boundary condition parameterized by $k$, in the large charge $Q$ limit.  This allows us to read off the coefficient $c_{3/2}$ as a function of $k$.  In \cite{Liu:2020uaz}, the AdS boson stars are classified to have $A_i$, $B_i$ and $C_i$ series of solutions.  The $A_i$ solutions are those that smoothly connect to the AdS vacuum as the charge $Q\rightarrow 0$.  The $B_i$ solutions, on the other hand, is gapped from the AdS vacuum, with the charge bounded below, but unbounded above.  In the $C_i$ solutions, the charge $Q$ is bounded both above and below.  All these solutions were known to exist previously in literature, e.g.~\cite{Gentle:2011kv}; however, their classification was first attempted in \cite{Liu:2020uaz} for the Dirichlet boundary condition $(k=\infty)$.  In particular, the $A_1$ series of boson stars, whose charges are in general unbounded above also, were identified as the ground states for a given charge $Q$, since they have the lowest (free) energy in the fixed charge ensemble \cite{Liu:2020uaz}. In this paper, we shall also focus on constructing the $A_1$ series of AdS boson stars. They can be easily identified since they have the smooth $Q\rightarrow 0$ limit, which excludes the $B_i$ and $C_i$ solutions.  The condition of lowest energy picks the $A_1$ out of the general $A_i$ series of boson stars.  We can then vary the integration constants and trace the solutions to those with large $Q$ and read off the $c_{3/2}$ coefficient.

The paper is organized as follows.  In section 2, we outline the setup and the numerical procedure to construct the AdS boson stars.  In section 3, we present the result of $c_{3/2}$, as a function of the boundary condition parameter $k$.  We conclude the paper in section 4.

\section{The setup}

{\bf Ansatz and equations}: In this section, we outline the numerical approach for constructing the boson stars.  We consider the spherically-symmetric and static metric with the ansatz
\bea
ds^2 &=& - h(r) dt^2 + \fft{dr^2}{f(r)} + r^2 d\Omega_2^2\,,\nn\\
A &=& a(r) dt\,,\qquad \Phi=\phi(r)\,.\label{genanz}
\eea
Here we have chosen the gauge where the phase $\chi$ of the complex scalar $\Phi=\phi e^{{\rm i} \chi}$ vanishes or ``is eaten'' by the vector $A_\mu$, which becomes massive.  The covariant equations of motion are
\bea
&& 2\Box\phi - A^2\fft{\partial U}{\partial\phi} - \fft{\partial V}{\partial\phi}=0\,,\qquad
\nabla_\mu F^{\mu\nu} -2 U A^\nu=0\,,\nn\\
&&R_{\mu\nu} -\ft12 g_{\mu\nu} R -3\ell^{-2} g_{\mu\nu}- \big(\partial_\mu\phi\partial_\nu\phi - \ft12 g_{\mu\nu} (\partial \phi)^2\big) - (A_\mu A_\nu - \ft12 g_{\mu\nu} A^2)\,U\nn\\
&&-\ft12 (F_{\rho\mu} F^{\rho}{}_\nu - \ft14 g_{\mu\nu} F^2) + \ft12 g_{\mu\nu}\, V=0\,,\label{coveom}
\eea
where $U=\ell^{-2} q^2 \phi^2 $ and $V=-2\ell^{-2}\phi^2 + \fft12\lambda \ell^{-2} \phi^4$. Substituting the ansatz (\ref{genanz}) into these covariant equations, we obtain differential equations for the functions $(f,h,a,\phi)$.  The function $f$ can be solved algebraically \cite{Liu:2020uaz}, and can be largely cast aside in the numerical approach. The remaining functions $(h,a,\phi)$ satisfy three coupled nonlinear second-order differential equations. This system  was well studied in literature and we shall not present here the detail different equations, which are cumbersome but straightforward to obtain.

{\bf Implementing the boundary condition}: The bulk action \eqref{genlag} alone is ill-defined in asymptotically AdS spacetime because it does not give rise to finite expressions for the variation of the action and
the on-shell values of the solutions. To cure this problem, certain counterterms must be included \cite{Balasubramanian:1999re,Bianchi:2001kw}. For the reduced system \eqref{genanz}, the total action in Lorentzian signature that allows the variational principle to be well-defined consists of the following pieces
\be
S_{\rm tot}=S_{\rm bulk}+S_{\rm GHY}+S_{\rm ct}\,,\label{Iren}
\ee
where the bulk action is given in \eqref{genlag}. The surface Gibbon-Hawking-York \cite{York:1972sj,Gibbons:1976ue} term and boundary counterterm are\footnote{For Dirichlet boundary condition with $\phi_1=0$, the variational principle is satisfied without introducing scalar dependent counterterms. However, for non-vanishing fixed $\phi_1$, these counterterms seem indispensable.}
\bea
S_{\rm GHY} &=&
 \frac{1}{8\pi G}\int_{\partial {\cal M}}d^3x \sqrt{-h}K\,,\nn\\
S_{\rm ct}&=&\frac{1}{8\pi G }\int_{\partial {\cal M}} d^3x
\sqrt{-h}\Big(-\frac2{\ell}-\frac{\ell}2{\cal R}
+(1-\alpha) \phi n^\mu\del_\mu\phi +
\fft{1-2\alpha}{2\ell}\, \phi^2 +
\fft{2\beta}{3\ell}\, \phi^3 \Big)\,,
\eea
where $(\alpha,\beta)$ parameterize the degree of freedom for which the variation of the action is finite. ({\it A priori}, the three scalar terms in $S_{\rm ct}$ would have three coefficients, instead of the two.)  The covariant expression for the scalar dependent counterterms specific to the model considered here was first given in \cite{Cremonini:2014gia}. In the expressions above, $K$ is the trace of the extrinsic curvature of the boundary surface, with $n^{\mu}$ being the outward unit normal vector, and ${\cal R}$ is the Ricci scalar associated with the induced boundary metric $h_{\mu\nu}$. Using \eqref{Iren} and substituting the Fefferman-Graham expansion, we can check that the variation of $\phi$ yields the finite boundary term
\be
\delta_{\phi}S_{\rm tot}\propto \int_{\partial {\cal M}} d^3x \left({\alpha}\phi_2\delta\phi_1
+(\alpha-1)\phi_1\delta\phi_2+2\beta\phi_1^2\delta\phi_1\right)\,,
\ee
which vanishes upon $\phi_2=k\phi_1^2$, provided that
\be
(\ft{3}2 \alpha-1)k+\beta=0\,.
\ee
For $\phi_2-k\phi_1^2=J$ with fixed $J$, only the choice $\alpha=0\,,\beta=k$ gives the desired form
\be
\delta_{\phi}S_{\rm tot}=-\frac1{8\pi G} \int_{\partial {\cal M}} d^3x\, \phi_1\delta\left(\phi_2-k\phi_1^2\right),
\ee
meaning that indeed, $\phi_1$ corresponds to the VEV while the combination $\phi_2-k\phi_1^2$ is associated with the source, in the same spirit of \cite{Witten:2001ua}.

{\bf Asymptotic Structure}: The Lagrangian admits the AdS vacuum of radius $\ell$, given by
\be
ds_{\rm AdS}^2 = -(\ell^{-2} r^2 + 1) dt^2 + \fft{dr^2}{\ell^{-2} r^2 + 1} + r^2 (d\theta^2 + \sin^2\theta\, d\varphi^2)\,,\qquad \phi=0=a'.\label{adsvac}
\ee
Since the scalar $\phi$ is conformally massless, the leading asymptotic boundary falloffs are
\be
\phi=\fft{\phi_1}{r} + \fft{\phi_2}{r^2} + \cdots\,.\label{scalarhair}
\ee
Their contributions to the leading asymptotic falloffs of the metric and the vector are given by
\bea
h &=&  \ell^{-2} r^2 + 1 - \fft{2M}{r} + \fft{8Q^2 + \ell^2 \phi_1^2 q^2 \mu^2}{2r^2} + \cdots\,,\nn\\
f &= & \ell^{-2} r^2 + 1 + \fft{\phi_1^2}{2\ell^{2}}- \fft{6\ell^2 M + 4 \phi_1\phi_2}{3\ell^2 r}
+ \fft{\ell^2(8Q^2 + \phi_1^2) + \phi_1^4 + 2 \phi_2^2}{2\ell^2 r^2} + \cdots\,,\nn\\
a &=&  \mu - \fft{4Q}{r} + \fft{\ell^2\phi_1^2 q^2\mu}{r^2} + \cdots \,.\label{falloffs}
\eea
Note that the parameter $\lambda$ does not enter the leading asymptotic falloffs. The asymptotic structure is characterised by five integration constants $(M,Q, \mu, \phi_1,\phi_2)$, which are the mass, electric charge, chemical potential and two scalar hair parameters respectively. It should be mentioned here that we consider only $k\ge 0$.  This is because when $k$ is negative, we can map it to the positive value by $\phi\rightarrow -\phi$, which is the symmetry of our system.

Using \eqref{Iren}, we find that the trace of the holographic stress tensor and the energy density are given by \cite{Cremonini:2014gia}
\bea
\tau^\mu{}_\mu &=&\frac1{4\pi G}\left((2\alpha-\ft43) {\ell}^{-2} \phi_1\phi_2 +\ft43 {\ell}^{-2} \beta
\phi_1^3\right),\nn\\
{\cal E} &=& \frac1{8\pi G}\left(M + (\ft43-2\alpha) {\ell}^{-2} \phi_1\phi_2 - \ft43 {\ell}^{-2}\beta
\phi_1^3\right).
\eea
From the expressions above, we see that for
\be
\phi_2=k\phi_1^2\,,\quad (\ft{3}2 \alpha-1)k+\beta=0\,,
\ee
the holographic stress tensor is traceless implying that the boundary condition indeed preserves the conformal symmetry in the large central charge limit. Meanwhile, the energy of the solution is still given by the parameter $M$ without explicit scalar-dependent contribution. For the mixed boundary condition \eqref{bc1}, the energy computed this way equals to the one obtained by integrating the infinitesimal conserved charge associated with Killing vector $\partial_t$ \cite{Lu:2013ura}. The electric charge is given by
\be
Q=\frac1{16\pi}\int *F\,.
\ee

In our system, the scalar can be set zero consistently, in which case, the theory admits the Reissner-Nordstr\"om (RN) AdS black hole. In the extremal limit, the mass becomes a function of charge $Q$, and the mass/charge relation can be analytically obtained, giving, for $\ell=1$, in the factorized form:
\be
M = \frac{\sqrt{\sqrt{48 Q^2+1}-1} \left(\sqrt{48 Q^2+1}+2\right)}{3 \sqrt{6}}\,.\label{rn-ads}
\ee
 By the AdS/CFT correspondence and state operator correspondence, asymptotically AdS solutions are dual to heavy operators whose scaling dimensions are related to the mass of bulk solution by $M=\frac{32 \Delta(Q)}{\pi c_T}$ \cite{Loukas:2018zjh,Liu:2020uaz}, where $c_T$ is the central charge of the dual CFT. In the large $Q$ expansion, one has $M\sim Q^{3/2}$ with the proportionality constant given by
\be
c_{3/2}^{\rm RN}=\frac{4 \sqrt{2}}{3^{3/4}}\sim 2.482\,.
\ee

{\bf The near origin expansion}: We are interested in AdS boson star solutions that are solitons with no curvature singularity. They are spherically symmetric and asymptotic to the AdS spacetime. The origin $(r=0)$ of a boson star is the Minkowski spacetime, arising from the ``boundary'' condition
\be
f=1\,,\qquad h=h_0\,,\qquad a=a_0\,,\qquad \phi=\phi_0\,,\qquad f'=h'=a'=\phi'=0\,.\label{middlecond}
\ee
Specifically, the condition for the geodesic completeness, or the absent of curvature singularity at $r=0$ requires that $h'=0=f'$. The equations of motion then dictate that $a'=0=\phi'$. However, we cannot use this boundary condition to solve the system numerically, since the differential equations become apparently singular at $r=0$.  This is because the general (singular) solutions at $r=0$  involve five nontrivial parameters. However, the $r=0$ condition (\ref{middlecond}) does give rise to consistent solutions at least locally and this can be demonstrated by performing the Taylor expansions of the boson star at the origin, namely
\bea
&& h=h_0(1 + h_2 r^2 + h_4 r^4 + \cdots)\,,\qquad
f=1 + f_2 r^2 + f_4 r^4 + \cdots\,,\nn\\
&& a=\sqrt{h_0} \Big(a_0 + \tilde a_2 r^2 + \tilde a_4 r^4 + \cdots\Big)\,,\qquad
\phi=\phi_0 + \tilde \phi_2 r^2 + \tilde \phi_4 r^4 + \cdots\,.
\eea
The coefficient $h_0$ represents the time scaling invariance, which must be fixed appropriately so that the metric becomes the standard AdS (\ref{adsvac}) asymptotically. The higher order coefficients can {\it all} be solved in terms of $(\phi_0,a_0)$, {\it e.g.}
\bea
h_2 &=& \ell^{-2} (1 + \ft13 \phi_0^2 -\ft1{12} \lambda \phi_0^4) + \ft1{3} q^2 a_0^2\phi_0^2 \,,\qquad
f_2= \ell^{-2} (1 + \ft13 \phi_0^2 -\ft1{12} \lambda \phi_0^4) - \ft1{6} q^2 a_0^2\phi_0^2 \,,\,,\nn\\
\tilde a_2 &=&\ft13 q^2 a_0 \phi_0^2\,,\qquad\tilde \phi_2 = \ell^{-2} (-\ft13\phi_0^2 + \ft16 \lambda \phi_0^3) -\ft16 q^2 a_0^2\phi_0\,.
\eea
One can then perform numerical analysis to show that for appropriate $(a_0,\phi_0)$ at $r=0$, one can indeed integrate out to asymptotic AdS infinity. It follows that the general boson stars involve two parameters, leading to the generic asymptotic scalar hair $(\phi_1,\phi_2)$, defined by (\ref{scalarhair}). We can choose general mixed boundary condition (\ref{bc1}) that preserve the conformal symmetry. Thus for given fixed $k$, the boson star involve only one parameter and we can choose this parameter to be the electric charge $Q$. The other asymptotic parameters, such as the mass $M$ and the chemical potential $\mu$ are then functions of the charge.  We are interested in particular the mass/charge relation, which in the large $Q$ limit, takes the form
\be
M(Q,k)= c_{3/2} (k) Q^{3/2} + c_{1/2}(k) Q^{1/2} + \cdots\,.
\ee
(This structure was generally argued in \cite{Liu:2020uaz}.)  The focus of this paper is to obtain how the leading coefficient $c_{3/2}$ depends on the mixed boundary condition parameter $k$.

{\bf The Numerical approach}: To perform the numerical analysis, we can use the Taylor expansion as the correct initial values so that we can integrate out not literally from $r=0$, but slightly away from the origin such as $r_i=10^{-3} \ell$.  In order to maintain higher accuracy, we perform the Taylor expansions up to and including the tenth order of $r$.   Using the Taylor expansion as our initial data, we can integrate the solution from $r_i$ to the asymptotic infinity, or practically, some large $r_f$ value and then perform the curve fitting of the numerical results with the asymptotic structure (\ref{scalarhair}) and (\ref{falloffs}).  We can thus read off the parameters $(M,Q,\mu,\phi_1,\phi_2)$ from
the asymptotic structure of the functions $(h,a,\phi)$. (The function $f$ can be solved algebraically from $(h,a,\phi)$ and their derivatives and therefore out of the numerical consideration.) The $1/r^2$ terms of functions $(h,a)$ in (\ref{falloffs}) provide consistency check. We also have to organize the solutions by the asymptotic mixed boundary condition parameter $k$. By scanning the two-dimensional plane $(\phi_0,a_0)$, we obtain, for each given $k$, one parameter family of boson stars, parameterized by the electric charge $Q$.

In \cite{Liu:2020uaz}, boson stars with the Dirichlet boundary condition $(k=\infty)$ were classified by the scalar hair properties through scanning the $(\phi_0,a_0)$-parameter plan that defines the middle spacetime.  Three classes of solutions, named $A_i$, $B_i$ and $C_i$, solutions emerge. The $A_i$ boson stars connect smoothly to the AdS vacuum as $Q\rightarrow 0$.  The $B_i$ boson stars have a mass gap, with $Q$ bounded below, but unbounded above.  The $C_i$ boson stars carry the charge that is bounded both above and below.  In particular, the $A_1$ boson stars have the lowest energy for given charge, and therefore correspond to the ground states of the system for the fixed charge ensemble.  For the theory with no scalar self interaction $(\lambda=0)$, the $A_1$ boson star can also have charge $Q$ unbounded above for $q>q_c$ \cite{Gentle:2011kv}. In this paper, we find that the $A_1$ series of boson stars continue to exist in mixed boundary conditions. These boson stars are easiest to identify.  We can start with a sufficiently small $\phi_0$, corresponding to a boson star that is close to the AdS vacuum.  This excludes the $B_i$ and $C_i$ boson stars that have the mass gap from the vacuum. We can then scan $a_0$, and plot the function $(\phi_2-k \phi_1^2)$ as a function of $a_0$.  The roots, labelled as $i=1,2,\ldots,$ from small to large, give rise to the $A_i$ series of boson stars. As we increase $\phi_0$, the $a_0$ roots become larger and the charge $Q$ also become larger, but only the $A_1$ series can have charge unbounded above. We are interested in boson stars with large $Q$, we therefore construct the $A_1$ series, obtain the large $Q$ limit and read off the coefficient $c_{3/2}(k)$.

Our curve fitting analysis suggests that in the large $Q$ limit, the mass/charge relation is analogous to that of the RN-AdS black hole, taking the form
\be
M(Q,k) = c_{3/2}(k) Q^{3/2} + c_{1/2}(k) Q^{\fft12} + c_0 + c_{-1/2}(k) Q^{-1/2} + \cdots\,.
\ee
Numerical evidence suggests that
\be
\fft{c_{1/2}(k)}{c_{3/2}(k)}<1\,.
\ee
Thus for the charge $Q$ well over 1000, we can read off the coefficient with sufficient accuracy simply by taking approximately
\be
c_{3/2}(k) \sim \fft{M(Q,k)}{Q^{3/2}}\,.
\ee

\section{The results}

The main result of this paper is to obtain the coefficient $c_{3/2}$ as a function of $k$.  Having set the AdS radius $\ell=1$, our comformally massless scalar is characterized by the fundamental electric charge $q$ and the coupling $\lambda$ of the fourth-order self interaction.  For all the $A_1$ boson star solutions (with unbounded $Q$) we have constructed in this paper, we find that the numerical data can be well fit by the analytical expression
\be
c_{3/2}(k) = \fft{b_1 + b_2 k + b_3 k^{3/2} + b_4 k^2}{1 + b_5 k + b_6 k^{3/2} + b_7 k^2}\,,\label{ckgenfun}
\ee
where $(b_1, b_1, \ldots, b_7)$ are parameters that depend on $q$ and $\lambda$.\footnote{It follows from the first law of boson star dynamics $dM=\mu dQ$ \cite{Liu:2020uaz} that we have $\lim_{Q\rightarrow\infty} (\mu /Q^{1/2})=\fft32 c_{3/2}$.} The mixed boundary condition parameter runs from 0 to $\infty$. For small $k$, we have
\be
c_{\fft32}(k) = b_1 + (b_2 -b_1 b_5) k + (b_3 - b_1 b_6) k^{\fft32} + {\cal O}(k^2)\,.\label{smallk0}
\ee
For large $k$, we have
\be
c_{\fft32}(k) = \fft{b_4}{b_7} - \fft{b_4 b_6 - b_3 b_7}{b_7^2\,\sqrt{k}} + {\cal O}\Big({\fft1{k}}\Big)\,.\label{largek0}
\ee
In other words, the coefficient $c_{3/2}$ is linearly dependent on $k$ for sufficiently small $k$, and it falls off as $-1/\sqrt{k}$ for large $k$.  In all our examples, we find that $b_i$ coefficients are positive and therefore there is no singularity for $c_{3/2}$ at any finite $k$.  Furthermore, the function $c_{3/2}(k)$ is a monotonously increasing function of $k$ with no extremum at any finite $k$.

Explicitly, we first present the case where the scalar has no self interaction, namely $\lambda=0$.  The existence of the $A_1$ boson stars requires that the parameter $q$ be larger than some critical value $q_c$, which was established to be $q_c^2=1.259$ in \cite{Gentle:2011kv} and $q_c^2=1.261$ in \cite{Liu:2020uaz}.  Here, we present the results for the $q^2=1.3, 1.4 $ and $1.5$ three cases. The explicit $b_i$ coefficients are summarized in Table.~1:
\be
\begin{tabular}{|c|c|c|c|c|c|c|c|}
  \hline
  $q^2$ & $b_0$ & $b_1$ & $b_2$ & $b_3$ & $b_4$ & $b_5$ & $b_6$ \\
  \hline
  $1.3$ & $1.2180$ & $4.5659$ & $1.4908$ & $3.8911$ & $3.3198$ & $1.3811$ & $1.7365$ \\
  \hline
  $1.4$ & $1.1818$ & $4.3277$ & $1.4407$ & $3.6999$ & $3.2261$ & $1.3941$ & $1.6730$ \\
  \hline
  $1.5$ & $1.1490$ & $4.1484$ & $1.4161$ & $3.5084$ & $3.1725$ & $1.4047$ & $1.6071$ \\
  \hline
\end{tabular}
\ee
\centerline{Table 1. \small The list of $b_i$ coefficients of the function $c_{3/2}(k)$ for $\lambda=0$.}

\begin{figure}[htp]
\begin{center}
\includegraphics[width=270pt]{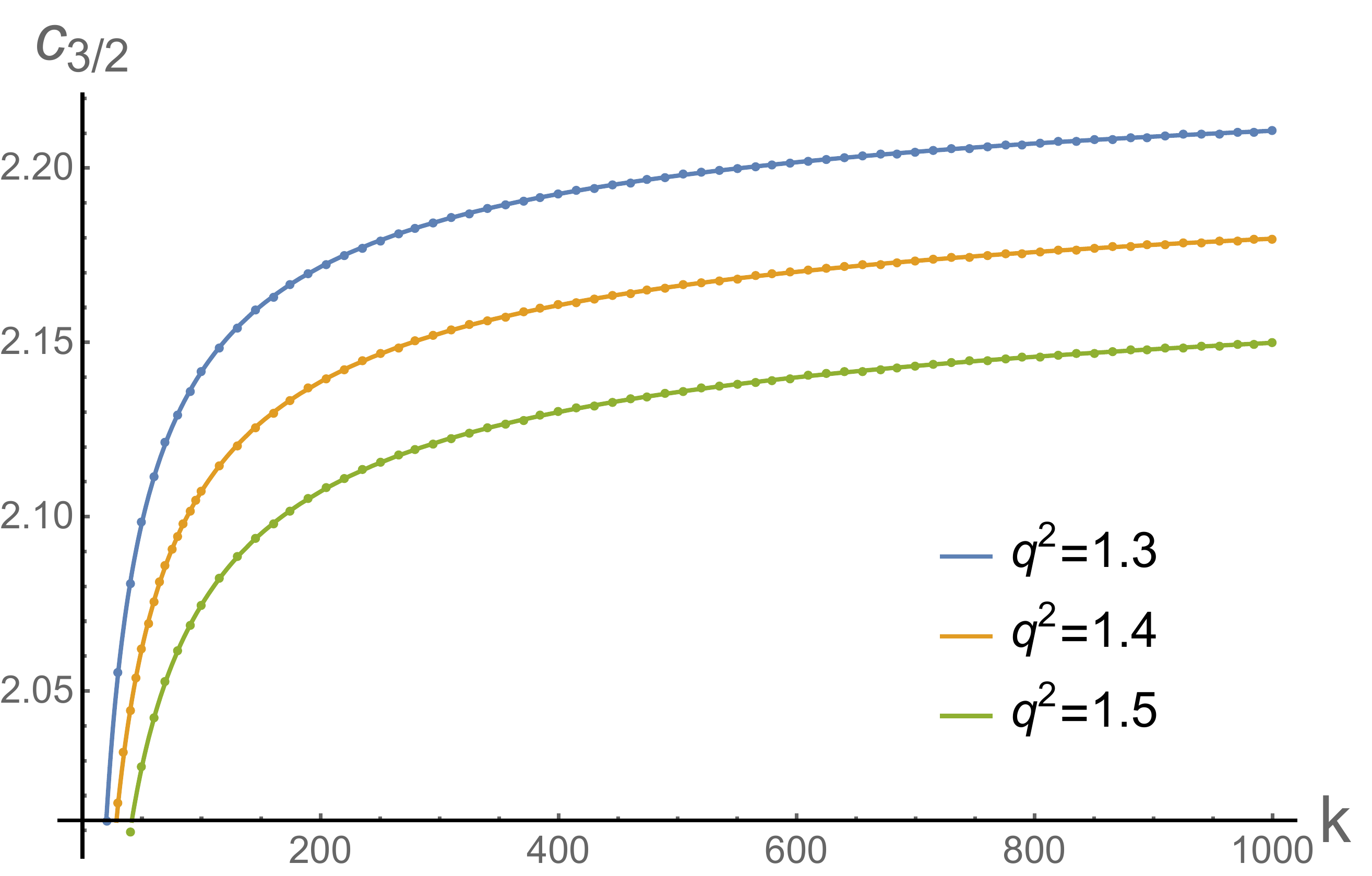}
\includegraphics[width=270pt]{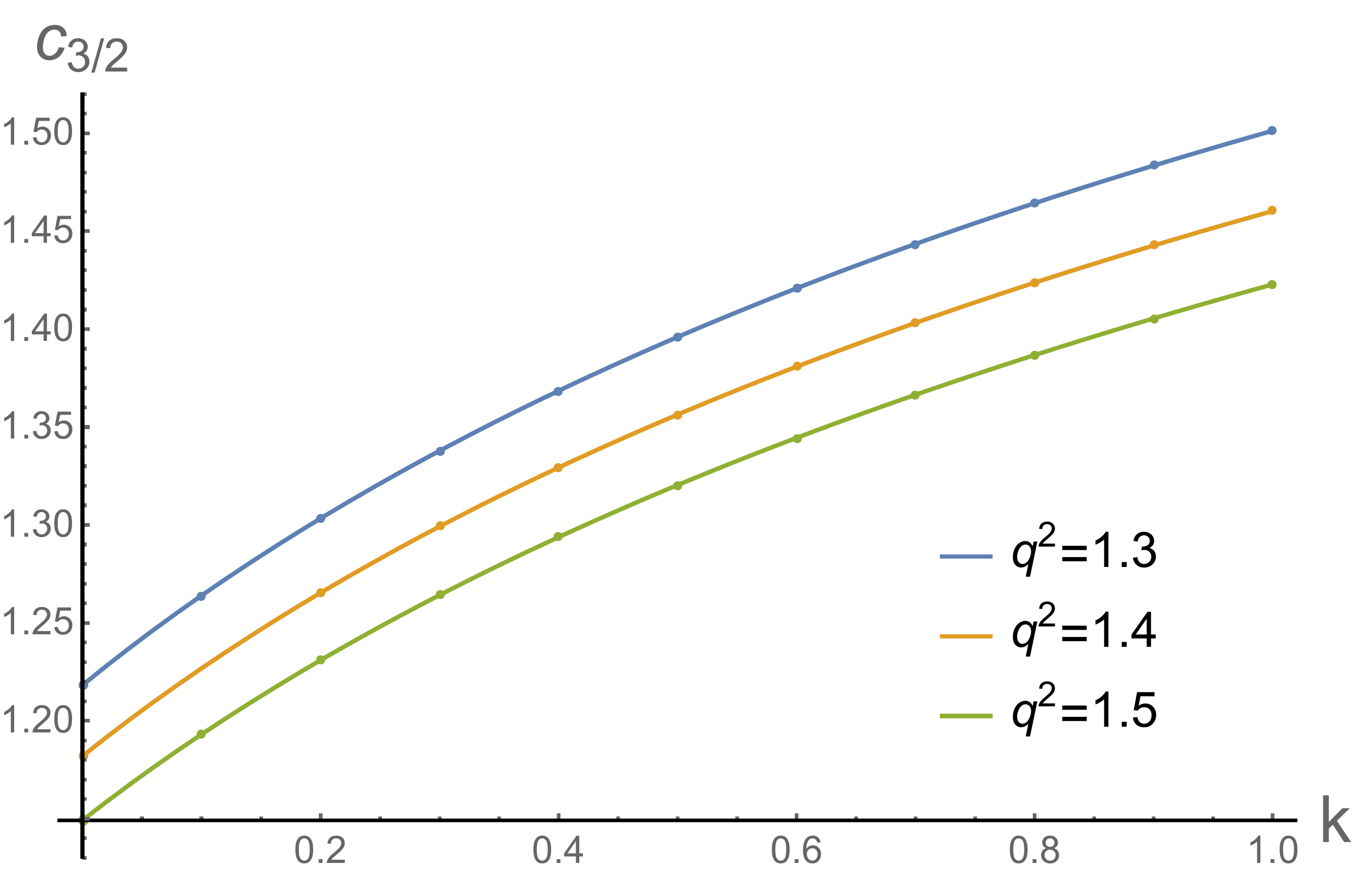}
\includegraphics[width=280pt]{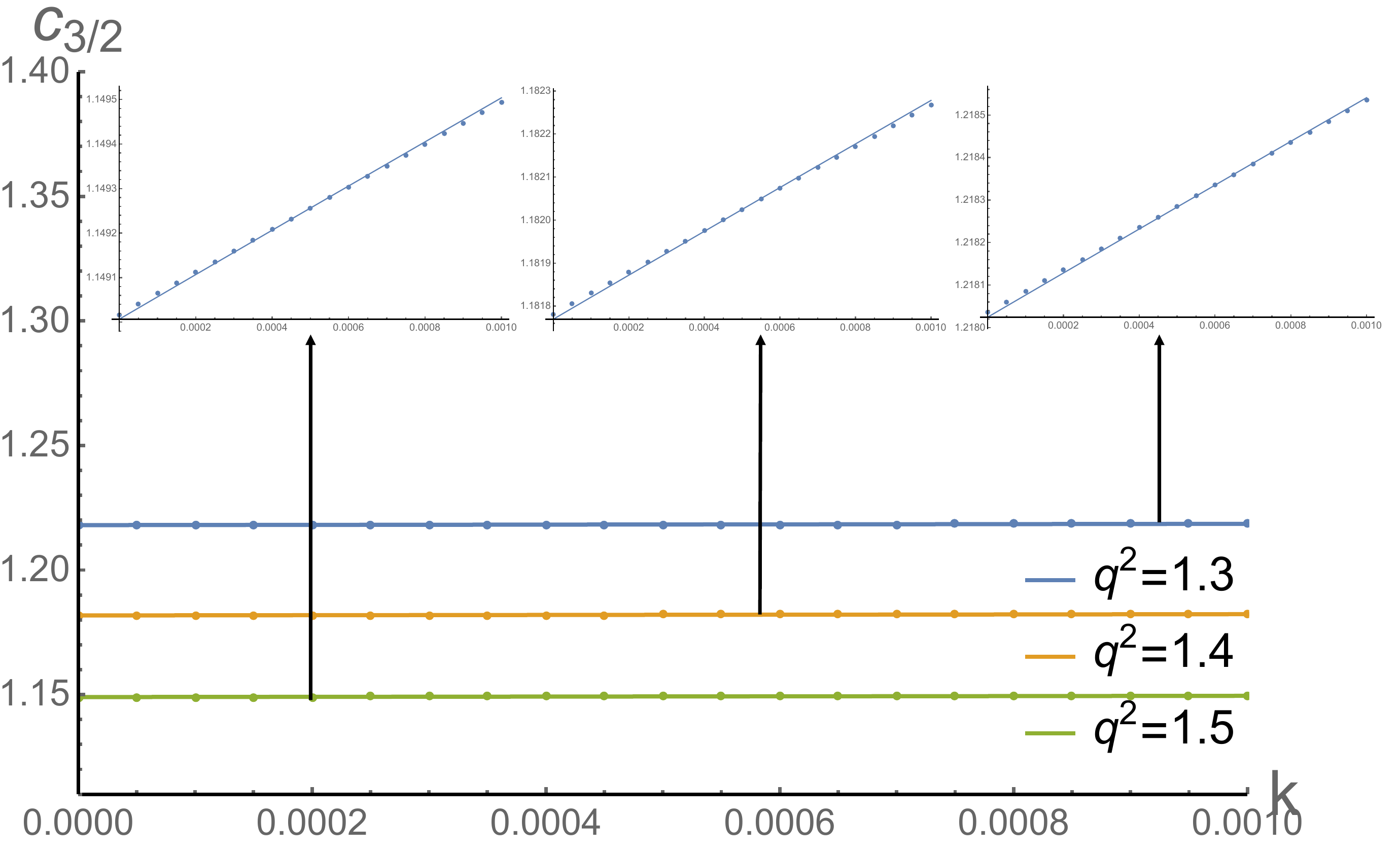}
\end{center}
\caption{\small\it The $c_{3/2}$ coefficient as a function of $k$ is plotted in the large, middle and small $k$ regions for different $q^2$.  The dots are the numerical data and lines are the analytical function (\ref{ckgenfun}), and they have perfect match even though the data runs over 7 orders of magnitude.  Note that the linear dependence at small $k$ appears to be horizontal when plotted together, actually the slops are all about $0.5$, as can be seen in the magnified plots.}
\label{q345-all}
\end{figure}

In order to illustrate how well the function (\ref{ckgenfun}) fits the numerical data, we plot them in Fig.~\ref{q345-all}.  We see that although our data cover a large range of $k$ of seven orders of magnitude, with minimum non-vanishing $k=10^{-4}$ to largest $k=1000$, our function (\ref{ckgenfun}) matches with the numerical data perfectly in the whole range.

For small and large $k$, the coefficient $c_{\fft32}$ behaves respectively as
\bea
q^2=1.3:&& c_{\fft32} \sim1.2180 + 0.52229 k - 0.19143 k^{3/2} \,,\qquad c_{\fft32}\sim 2.2407-0.92364 \sqrt{\frac{1}{k}}\,,\nn\\
q^2=1.4:&& c_{\fft32} \sim1.1818 + 0.51524 k - 0.20679 k^{3/2}\,,\qquad c_{\fft32}\sim 2.2116-0.98172 \sqrt{\frac{1}{k}} \,,\nn\\
q^2=1.5:&& c_{\fft32} \sim 1.1490 + 0.50318 k - 0.19786 k^{3/2}\,,\qquad c_{\fft32}\sim 2.1831-1.0270 \sqrt{\frac{1}{k}}\,.
\eea
In fact, the quantities $c_{3/2}(0)$ and $c_{3/2}(\infty)$ from the conjectured formula (\ref{ckgenfun}) also match the corresponding numerical data up to five significant figures. As the  $k^{3/2}$ term approaches 0 not much faster than $k$, in order to see the linearity at small $k$ clearly, the mixed boundary condition parameter $k$ has to be sufficiently small, less than $10^{-2}$.  Collecting the data for $0.1<k<1000$ is not enough for one to deduce the linear nature at small $k$, let alone the function (\ref{ckgenfun}).  The numerical data indicate that for $k\ll 1$, both $c_{3/2}$ and $\partial_k c_{3/2}$ decrease as $q$ increases. Correspondingly, the mass of the boson star at the large charge limit becomes smaller for fixed $Q$ as $q$ increases.

We also examine the effect of the self-interacting scalar term by turning on the $\lambda$. We obtain the results for $\lambda=0.1$ and 0.2 for fixed $q^2=1.4$.  The $b_i$ coefficients of the function (\ref{ckgenfun}) are summarized in Table 2:
\be
\begin{tabular}{|c|c|c|c|c|c|c|c|}
  \hline
  $\lambda $ & $b_1$ & $b_2$ & $b_3$ & $b_4$ & $b_5$ & $b_6$ & $b_7$ \\
  \hline
 $0.1$ & $1.2132$ & $4.6737$ & $1.5999$ & $3.9072$ & $3.4551$ & $1.4875$ & $1.7606$ \\
  \hline
  $0.2$ & $1.2414$ & $5.3812$ & $1.9227$ & $4.4974$ & $3.9660$ & $1.7289$ & $2.0198$ \\
\hline
\end{tabular}
\ee
\centerline{Table 2. \small The $b_i$ coefficients of the function $c_{3/2}(k)$ for $q^2=1.4$.}

\begin{figure}[htp]
\begin{center}
\includegraphics[width=220pt]{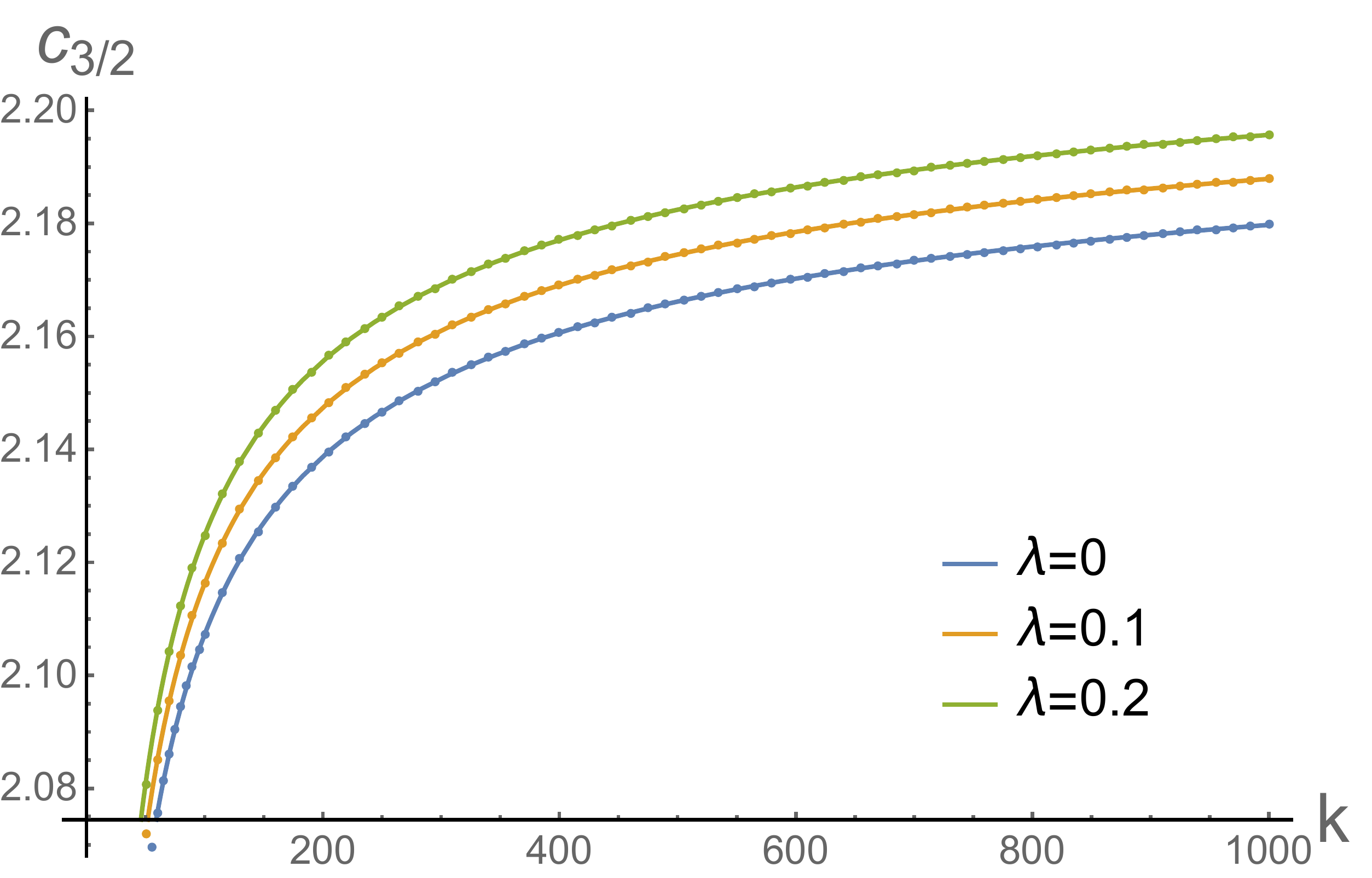}
\includegraphics[width=220pt]{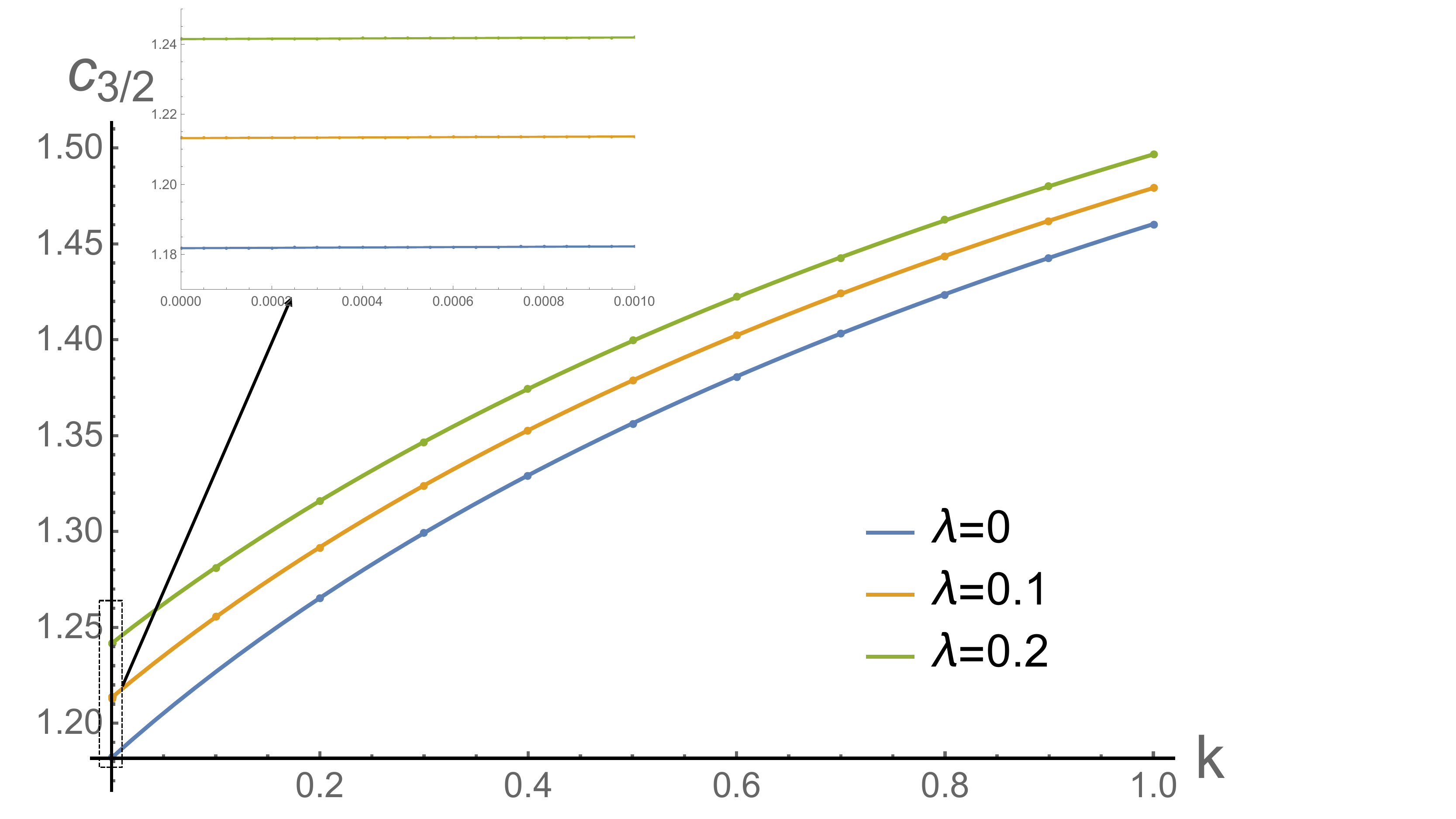}
\end{center}
\caption{\small\it The $c_{3/2}$ coefficient as a function of $k$ is plotted in the large, middle and small $k$ region for different $\lambda$, with fixed $q^2=1.4$.  The dots are the numerical data and lines are the analytical function (\ref{ckgenfun}), and they have perfect match even though the data run over 7 orders of magnitude.}
\label{lambda-all}
\end{figure}

For small and large $k$, the coefficient $c_{3/2}$ behaves respectively as
\bea
\lambda=0.1:&& c_{3/2} \sim 1.2132 + 0.48207 k - 0.20466 k^{3/2} \,,\qquad c_{\fft32}\sim 2.2192-0.96623 \sqrt{\frac{1}{k}}\,,\nn\\
\lambda=0.2:&& c_{3/2} \sim 1.2414 + 0.45768 k - 0.22353 k^{3/2}\,,\qquad c_{3/2}\sim 2.2266-0.95395 \sqrt{\frac{1}{k}} \,.
\eea

\begin{figure}[htp]
\begin{center}
\includegraphics[width=250pt]{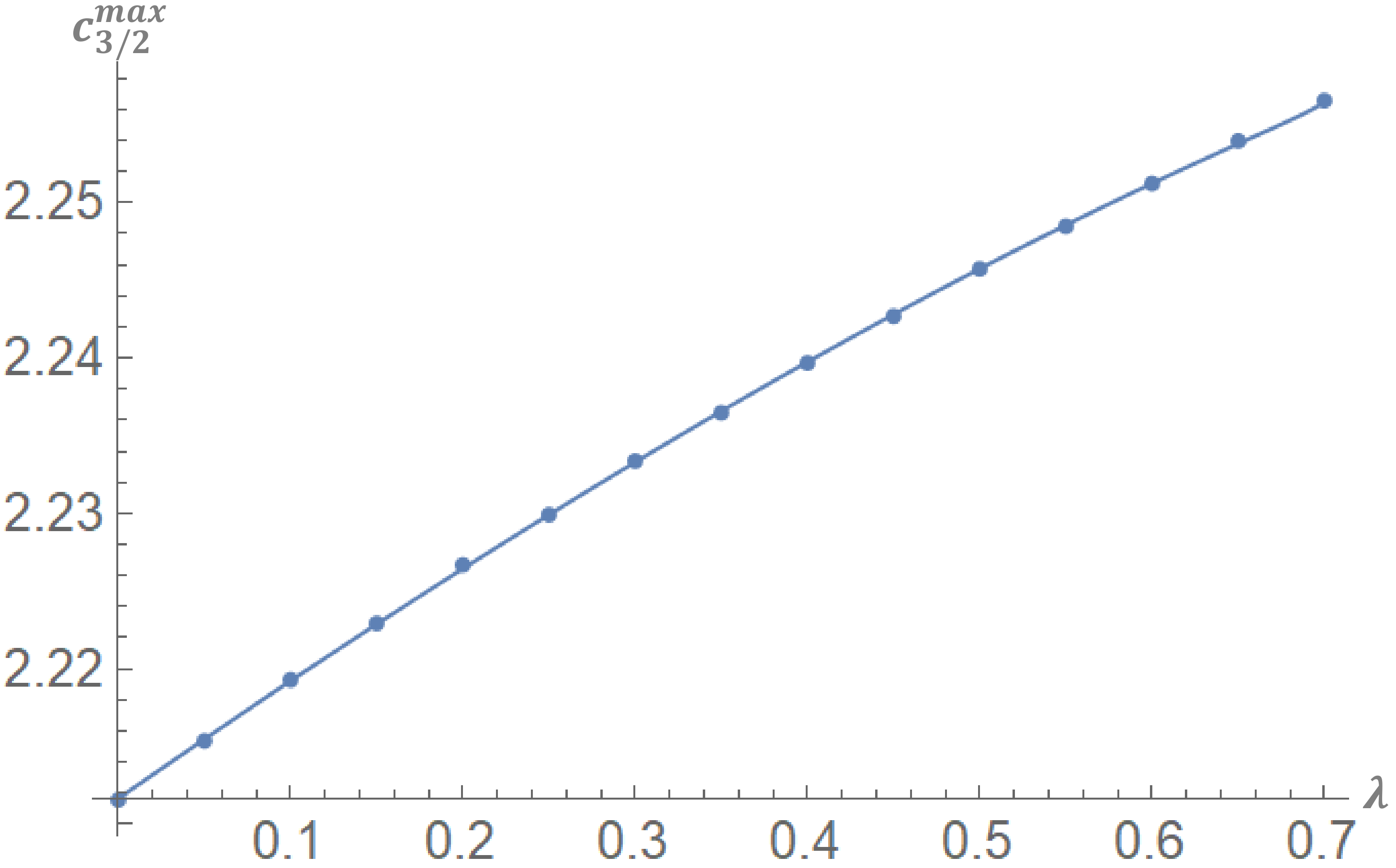}
\end{center}
\caption{\small\it The coefficient $c_{3/2}^{\rm max}\equiv c_{3/2}(\infty)$ increases as $\lambda$ becomes larger, but is always less than
$c_{3/2}^{\rm RN}$ when $\lambda\le \lambda^c\sim 0.70$, for which the charge $Q$ is unbounded above.}
\label{lambdauplimit}
\end{figure}

The result indicates that for fixed $k$, the coefficient $c_{3/2}(k)$ increases as $\lambda$ becomes larger.  This could lead a possibility that the coefficient becomes bigger than $c_{3/2}^{\rm RN}$ of the extremal RN-AdS black hole in which case the boson star no longer describes the ground state.  In order to study whether this could happen, we consider $c_{3/2}^{\rm max}$ and compare this to $c_{3/2}^{\rm RN}$.  The coefficient $c_{3/2}^{\rm max}$ arises when $k=\infty$, corresponding to $\phi_1=0$. We find that there is an upper bound for the parameter $\lambda$, namely $\lambda^c\sim 0.70$, beyond which, the charge of the $A_1$ series boson star becomes bounded above, analogous to the $A_1$ boson stars of the $SU(3)$ gauged supergravity model studied in \cite{Liu:2020uaz}.  For $\lambda\le \lambda^c$, the charge $Q$ is unbounded above and hence the coefficient $c_{3/2}$ is meaningful and calculable. We find that they are all less than $c_{3/2}^{\rm RN}$.  The $\lambda$ dependence is plotted in Fig.~\ref{lambdauplimit}.

Since $\lambda$ must be less than $\lambda_c$, we use leading-order polynomial expansions of $\sqrt{\lambda^c-\lambda}$ to fit the data.  This leads to
\bea
c_{3/2}^{\rm max} &=& 2.25662-0.00479295 \sqrt{0.70000-1.0000 \lambda}-0.026558 (0.70000-1.0000 \lambda)\nn\\
&&-0.0382011 (0.70000-1.0000 \lambda)^{3/2}.
\eea
As we can see from (\ref{lambdauplimit}) that it fits the data very well. However, since the parameter $\lambda$ lies in a small region and the function is smooth, any other non-singular expansion for the curve-fitting procedure could equally achieve a good fitting.

\begin{figure}[htp]
\begin{center}
\includegraphics[width=300pt]{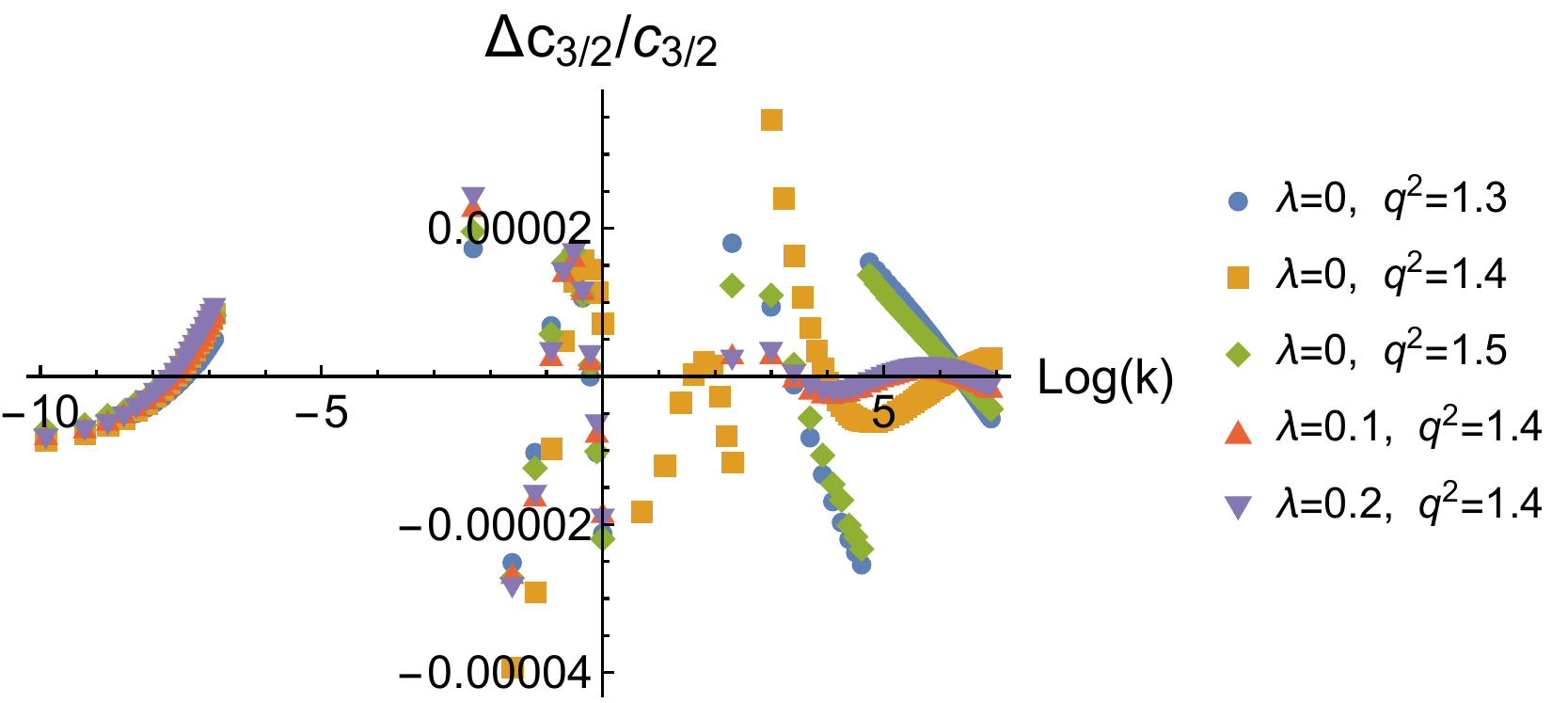}
\end{center}
\caption{\small\it Although $k$ runs over 7 orders of magnitude, the errors between the conjectured formula (\ref{ckgenfun}) and the numerical data are very small, with $|\Delta c_{3/2}|/c_{3/2}<4\times 10^{-5}$.
}
\label{error}
\end{figure}

Before ending this section, we would like to present the graph that shows the error of our formula (\ref{ckgenfun}) with the actual numerical data for minimum $k=1/20000$ to largest $k=1000$.  As we can see from Fig.~\ref{error} that the accuracy is high with $|\Delta c_{3/2}|/c_{3/2}<4\times 10^{-5}$.  The data for $k=0$ and $k=\infty$, which lie outside the graph, also have the same magnitude of accuracy.

Finally, we would like to give further comments on our conjectured seven-parameter expression (\ref{ckgenfun}) for the coefficient $c_{3/2}(k)$. The linear dependence (\ref{smallk0}) on $k$ for small $k$ is not surprising and can be easily established for sufficiently small $k$. The finite value of $c_{3/2}(k)$ at $k=\infty$ established previously in literature suggested Laurent-like expansion for large $k$.  However, the $-1/\sqrt{k}$ leading falloff (\ref{largek0}) could not be trivially prevised, but was determined by numerical data.  A rational polynomial is a natural and the simplest way to fit data for both small and large $k$ values. It turns out that our simple conjecture fits the numerical data with high accuracy for all $0\le k \le \infty$ values. The expression may give an impression that these are simply the leading order terms of rational power expansion, where the numerator and the denominator can be further augmented with higher power terms of $k$. This is in fact not the case. We considered adding higher order terms of $k^{5/2}, k^3$ with more non-vanishing parameters in both numerator and denominators and we found that curve fitting would always lead to a singular function. At this stage, the physical or mathematical significance of the fitting expression is unclear and deserves further study in the future.

\section{Conclusions}

In this paper, we studied AdS boson stars in the large charge limit with mixed boundary condition. The bulk theory is Einstein-Maxwell theory with a cosmological constant, coupled to a charged conformally massless complex scalar with fourth-order self-interaction, parameterized by the dimensionless parameters $q$ and $\lambda$ respectively. We focus on the $A_1$ series of boson stars that smoothly connect the AdS vacuum as $Q\rightarrow 0$ and also allow arbitrarily large $Q$. We obtain the coefficient $c_{3/2} = \lim_{Q\rightarrow \infty} (M/Q^{3/2})$, as a function of  parameter $k$, appearing in the mixed boundary condition defined by (\ref{bc1}).  We find that for proper values of $(q,\lambda)$ such that  the $A_1$ series of boson stars exists, the coefficient $c_{3/2}$ has the following properties
\begin{itemize}

\item $c_{3/2}(k)$ is a monotonously increasing function interpolating between the Neumann and Dirichlet values, i.e. with $c_{3/2}^{\rm N}=c_{3/2}(0)$ and $c_{3/2}^{\rm D}=c_{3/2}(\infty)$.

\item $c_{3/2}(k)$ grows linearly in $k$ for sufficiently small $k$, {\it i.e.}~$c_{3/2}(k)-c_{3/2}^{\rm min}\sim k$; it falls off as $c_{3/2}(k)-c_{3/2}^{\rm max}\sim -1/\sqrt{k}$ as $k\rightarrow \infty$.

\end{itemize}
Based on the numerical results, we proposed an analytical expression (\ref{ckgenfun}) involving seven parameters $(b_1,b_2, \ldots, b_7)$ that could be determined by curve fitting.  We found that the formula matched the numerical data with a striking accuracy of $|\Delta c_{3/2}|/c_{3/2}<4\times 10^{-5}$, for all $k$ running from zero to infinity. It should be pointed out that there is no reason to believe that (\ref{ckgenfun}) is the exact expression of $c_{3/2}(k)$; nevertheless, the high accuracy suggests that the formula captures the essence of its mixed boundary dependence.

As for future directions, it is certainly interesting to construct AdS boson stars with mixed boundary condition in higher dimensions, generalising the known solutions with Dirichlet boundary conditions  \cite{Basu:2010uz,Bhattacharyya:2010yg,Dias:2011tj}. The $D=5$ bulk results \cite{Basu:2010uz,Bhattacharyya:2010yg,Dias:2011tj} indicate that in the large $Q$ limit,
the mass of the AdS boson stars behave as $M(Q)\sim Q^{4/3}$ which also occurred in the lowest scaling dimension 
of charged operators in the critical $O(N$) model \cite{Antipin:2020abu} and  the walking $U(N)\times U(N)$ Higgs theory \cite{Antipin:2020rdw} in $4-\epsilon$ dimensions.
 Another important question is the stability of the solution. Amongst spherically symmetric solutions with $R\times S^2$ topology, we have seen that the boson star solutions with mixed boundary remains the ground state at a fixed $U(1)$ charge. Therefore, intuitively, they should be stable  against small perturbations. Previous studies \cite{Buchel:2013uba, Dias:2012tq} had provided strong evidence that   with Dirichlet boundary conditions, boson stars asymptotic to global AdS were stable for generic choices of initial values. It would be very interesting to extend their analysis to boson stars obeying mixed boundary conditions in the future.

\section*{Acknowledgement}

We are grateful to Zhan-feng Mai for useful discussions. This work is supported in part by NSFC (National Natural Science Foundation of China) Grants No.~12075166, No.~11675144, No.~11875200 and No.~11935009.

\end{document}